\documentclass[prd,showpacs,preprint,nofootinbib]{revtex4}
\usepackage{graphicx,xcolor,amsmath}
\usepackage{bm}

\usepackage{slashed}
\usepackage{extarrows}
\allowdisplaybreaks
\newcommand{\D}{{\rm{d}}}

\begin{document}
\title{The $H\rightarrow b\bar{s}$ decay and its implication for the
vector-like singlet fermion model}
\author{Jin Zhang\footnote{jinzhang@yxnu.edu.cn}}
\affiliation{Department of Physics, Yuxi Normal University,
Yuxi, Yunnan, 653100, China}
\author{Hong-Ying Jin\footnote{jinhongying@zju.edu.cn}}
\affiliation{Institute of Modern Physics, School of Physics, Zhejiang University,
Hangzhou, Zhejiang, 310027, China}
\author{T.G. Steele\footnote{Tom.Steele@usask.ca}}
\affiliation{Department of Physics and Engineering Physics,
University of Saskatchewan, Saskatoon, SK, S7N~5E2, Canada}


\begin{abstract}
The vector-like quark model is one
of the extensions of the standard model (SM) of particle physics.
The simplest version of this model introduces
a vector-like singlet quark which can mix with SM quarks
and give rise to new contributions to the
flavor-changing decays of the Higgs boson.
In this work we first present a systematic analysis
of the branching ratios of the decays $H\rightarrow b\bar{s}, b\bar{d}$
at leading order in the standard model. Our results show
that it is challenging to observe these two modes
because of their small branching ratios.
Then augmenting the SM with a vector-like singlet top quark, assuming the top partner
only mixes with the top quark, complete one-loop contributions
are taken into account in the amplitudes.
Further results indicate that the branching ratios of the
decays $H\rightarrow b\bar{s}, b\bar{d}$ are
sensitive to the mass of the top partner $M_{T}$
and the mixing effects characterized by $\sin\theta_{L}$.
By tuning the values of $M_{T}$ and $\sin\theta_{L}$,
the branching ratios may rise to a level accessible to LHC experiments.
Combined with the branching ratios obtained
from a probabilistic model, the allowed areas in the
$M_{T}-\sin\theta_{L}$ plane are displayed. Tagging efficiencies
and feasibility for detecting $H\rightarrow b\bar{s}$ are
specifically discussed and we conclude that with large statistics
it is promising to discover the $H\rightarrow b\bar{s}$ decay at the LHC.
\end{abstract}


\maketitle
\newpage

\section{introduction}
The discovery of the long-awaited Higgs boson
of the Standard Model~(SM)~\cite{Englert:1964et, Higgs:1964pj,
Higgs:1966ev, Guralnik:1964eu, Kibble:1967sv} by the ATLAS
and by the CMS collaborations at the LHC in 2012~\cite{ATLAS:2012yve, CMS:2012qbp}
marked a milestone in particle physics. Since the first observation,
substantial experimental data has been accumulated on
various decays of the Higgs boson in Run \uppercase\expandafter{\romannumeral 1}
and Run \uppercase\expandafter{\romannumeral 2} as well as the ongoing
Run \uppercase\expandafter{\romannumeral 3} of the LHC.
Full study of the decays of the Higgs boson holds a prominent
role in deciphering physics of the SM. It is known that a Higgs boson
with mass about $125\,{\rm{GeV}}$ can decay to many particles in
the SM~\cite{PDG2022Higgs}.  We may classify the dominant decay
modes of the Higgs boson into two main categories according to
the final state particles. The first category is decays to vector bosons,
and the second category is decays to flavor-conserving fermion pairs
which can occur at tree level in the SM. Abundant events of these
two main categories have been observed. On the theoretical side,
these two categories of decay modes have been
evaluated to higher order in perturbation
theory in the SM and its supersymmetric extensions~\cite{Gunion:1989we,
Kniehl:1993ay, Spira:1997dg, Djouadi:2005gi, Djouadi:2005gj, Spira:2016ztx, Choi:2021nql}.
Since flavor-changing neutral currents~(FCNC)
are forbidden at tree level in the SM, at leading order
in perturbation theory, quark flavor changing decays of
the Higgs boson~(denoted by $H\rightarrow q\bar{q}'$)
are mediated via triangle diagrams. It has been widely
discussed that the investigation of quark flavor-changing
decays~\cite{Willey:1982mc, Grzadkowski:1983yp, Benitez-Guzman:2015ana,
Barducci:2017ioq, Aranda:2020tqw, Kamenik:2024mpe} of the
Higgs boson can offer practical clues for models of new physics
beyond the SM, such as two-Higgs-doublet models~\cite{Arhrib:2004xu, Barenboim:2015fya,
Crivellin:2017upt, Arco:2023hmz},
supersymmetric models~\cite{Bejar:2004rz, Hollik:2005as, Arhrib:2006vy,
Barenboim:2015fya, Gomez:2015duj}, extra dimensions~\cite{Farrera:2020bon} and
fourth generation models~\cite{Eilam:1989zm}.

However, observation of the quark flavor-changing decays
need high statistics. Taking the decay $H\rightarrow b\bar{s}$ as an example,
a qualitative estimate shows that it is more difficult
to detect it than any decay modes of the Higgs boson
observed at the LHC because the amplitude of
this decay is suppressed in several ways. Firstly,
at leading order in the SM, the process is mediated by
triangle diagrams, so the squared amplitude of the process is
suppressed by $G_{f}^{3}$, where $G_{f}$ denotes the Fermi constant.
On the other hand, the CKM elements~\cite{Cabibbo:1963yz, Kobayashi:1973fv}
will provide further suppression at the order-of-magnitude about $10^{-2}$ or less.
Finally, with the addition of heavy quarks, the contribution stemming
from the Higgs couplings to light quarks are so small that they can be neglected.
But thanks to the small width of the decay $H\rightarrow b\bar{s}$,
it provides an opportunity to explore effects of new physics
in that a small SM background give us the chance of observing
the small effects of new physics. In this sense, a comprehensive study
on the branching ratios of $H\rightarrow b\bar{s}$ and other
quark flavor changing Higgs decays can provide constraints
on new physics models in which new particles can contribute
to the amplitudes of the quark flavor changing Higgs decays.

In the quest for the new physics beyond the SM, the vector-like fermion models
have drawn much attention for decades~\cite{delAguila:1985mk, Branco:1986my,
Fishbane:1984zv, Fishbane:1985gu, Leike:1998wr, Hewett:1988xc,
Aguilar-Saavedra:2002phh, Kang:2007ib, Cacciapaglia:2011fx,
Cacciapaglia:2018qep, Ellis:2014dza, Arhrib:2016rlj,
Cacciapaglia:2017gzh, Kim:2018mks, Alhazmi:2018whk, He:2020suf,
Darvishi:2016gvm, Arsenault:2022xty, Corcella:2021mdl, Okada:2012gy, Deandrea:2021vje,
Carvalho:2018jkq}~(see Ref.~\cite{Alves:2023ufm} and references therein
for an up-to-date review on this subject). A practical version of these
models may introduce a new $U'(1)$ gauge group~\cite{Kamenik:2017tnu},
which is spontaneously broken by the vacuum expectation
of a scalar field $\Phi$, transforming as $\Phi\sim(1, 1, 0, q')$
under $SU_{C}(3)\times SU_{L}(2)\times U_{Y}(1)\times U'(1)$.
The model contains a colored Dirac fermion transforming as $T'\sim(3, 1, 2/3, q')$
which is often referred to as a top partner. The top partner
also arises in little Higgs models~\cite{Arkani-Hamed:2002ikv,
Low:2002ws, Perelstein:2003wd, Chang:2003un, Chen:2003fm, Han:2005ru, Hubisz:2005tx},
topcolor models~\cite{Hill:1991at, Hill:2002ap},
and top condensate models~\cite{Dobrescu:1997nm, Chivukula:1998wd,He:1999vp, Fukano:2012qx}.
In principle, the top partner will mix with SM quarks, but the mixing with the
first two generations is highly restricted by precision electroweak data
and flavor-changing neutral processes at low energies~\cite{Aguilar-Saavedra:2002phh}.
Thus it is reasonable to assume that the top partner only mixes with the top quark.
In generalized versions of these models, the vector-like fermions may be
a $SU_{L}(2)$ doublet or triplet~\cite{Cacciapaglia:2018lld, Dawson:2012di,
Aguilar-Saavedra:2013qpa, Chen:2017hak}. The top partner,
regardless of being in singlet or in the generalized models,
will introduce new contributions to the amplitude of $H\rightarrow b\bar{s}$,
thereby altering the decay width at a level which may be
accessible to LHC experiments. Therefore, a careful analysis
on the $H\rightarrow b\bar{s}$ decay may provide an alternative
approach to constraining the vector-like singlet model other
than trying to test it through the top partner
decays to SM particles.\footnote{See the summary
tables in Ref.~\cite{Alves:2023ufm} for an exhaustive
compilation of searches for vector-like singlet quarks by ATLAS and by CMS.}

Having noted  the small SM $H\rightarrow b\bar{s}$
width and the associated potential for testing
the vector-like singlet quark model by this decay process,
in this paper we will present a systematic analysis of the
$H\rightarrow b\bar{s}$ decay at leading order in SM and then
in the vector-like singlet top partner model.\footnote{The
$H\rightarrow b\bar{s}$ decay has been analyzed in Ref.~\cite{Kamenik:2024mpe}
based on the model of a single generation of vector-like singlet down-type quarks.
One can refer to the supplementary materials of Ref.~\cite{Kamenik:2024mpe} for details.}
We firstly evaluate the widths and branching fractions in the SM,
then the contribution of top partner to the amplitudes will be considered.
We assume that the top partner only mixes with the top quark,
and mixing with the first two generation quarks is neglected.
Another down-type flavor-changing decay of the Higgs boson,
$H\rightarrow b\bar{d}$, can also be analyzed by appropriate
replacement of the parameters in the corresponding expressions.

The paper is organized as follows. In Section~\ref{SM_section}
the branching ratios of the processes $H\rightarrow b\bar{s}, b\bar{d}$
in the SM are evaluated, and the analytic and
numerical results are presented. In Section~\ref{BR_section}
the contributions of the top partner are taken into account,
and the role of top partner mass and mixing effects are investigated.
Employing the upper bounds of the branching ratios of $H\rightarrow b\bar{s}$
from experiments and using a probabilistic model, the allowed parameter spaces are obtained.
The tagging efficiencies and detection feasibility of the
decay $H\rightarrow b\bar{s}$ at the LHC are discussed.
Our conclusions and outlooks are summarized in Section~\ref{summary_section}.
Some necessary formulae are collected in the Appendices.

\section{evaluation the decay rate of $H\rightarrow b\bar{s}$ in the Standard Model}
\label{SM_section}
\subsection{the amplitude formulae}

\begin{figure}
\begin{center}
\includegraphics[scale=0.50]{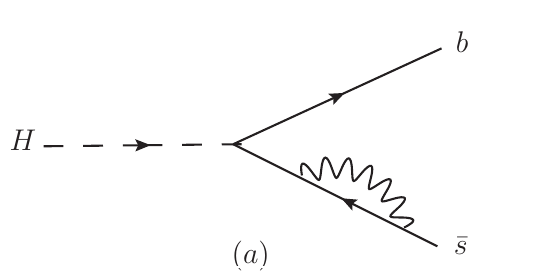}
\includegraphics[scale=0.50]{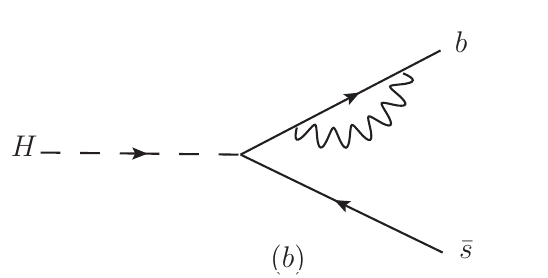}\\
\includegraphics[scale=0.50]{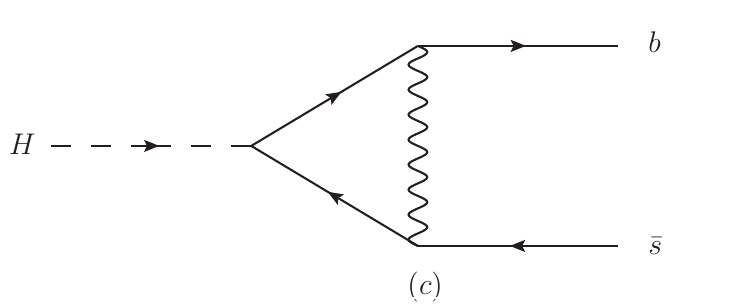}
\includegraphics[scale=0.50]{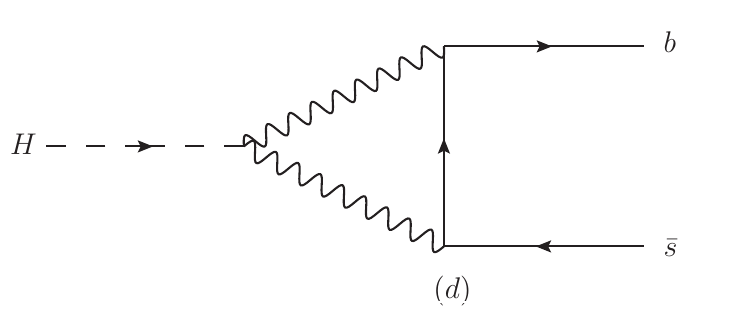}
\caption{One-loop diagrams contributing to the
$H\rightarrow b\bar{s}$ decay in the SM.} \label{twotypediagofthestandardmodel}
\end{center}
\end{figure}

All the one-loop diagrams\footnote{The Feynman diagrams in this paper
are produced by \textsf{Jaxodraw}~\cite{Binosi:2003yf, Binosi:2008ig}.}
contributing to the $H\rightarrow b\bar{s}$ decay
in the SM are depicted in Fig.~\ref{twotypediagofthestandardmodel}.
We may divide them into two groups: diagrams (a) and (b) represent
quark self-energy corrections and diagrams (c) and (d) represent
triangle diagrams. The amplitude of diagram (a) in
Fig.~\ref{twotypediagofthestandardmodel} in Feynman-'t Hooft gauge is
\begin{eqnarray}
M_{a}&=&-\frac{m_{b}(\sqrt{2}G_{f})^{1/2}g_{2}^{2}}{2}
\sum_{q}V_{qb}^{*}V_{qs}\int\frac{\D ^{4}k}{(2\pi)^{4}}
\frac{1}{D_{1}D_{2}D_{3}}\nonumber\\
&\times&\bar{u}(p_{2})(\slashed{p}_{3}+m_{b})\gamma_{\mu}P_{L}
(\slashed{p}_{3}-\slashed{k}+m_{q})\gamma^{\mu}P_{L}v(p_{3}), \label{gourponediagrama}
\end{eqnarray}
where the summation over $q=u,c,t$ is implied, the denominators are
\begin{equation}
D_{1}=p_{3}^{2}-m_{b}^{2}+i\varepsilon,\quad
D_{2}=k^{2}-m_{W}^{2}+i\varepsilon,\quad
D_{3}=(k-p_{3})^{2}-m_{q}^{2}+i\varepsilon,    \label{thrdenominatorsofdiagramagroupone}
\end{equation}
$V^{*}_{qb}$ and $V_{qs}$ are the elements of
the CKM matrix~\cite{Cabibbo:1963yz, Kobayashi:1973fv}, and
$G_{f}$ is the Fermi coupling constant~\cite{PDG2022Higgs}
\begin{equation}
G_{f}=\frac{g_{2}^{2}}{4\sqrt{2}m_{W}^{2}}
=1.1664\times10^{-5}\,{\rm{GeV}}^{-2}. \label{fermiconstant}
\end{equation}
We denote the momentum of the Higgs boson,
$b$ and $\bar{s}$ to be $p_{1}$, $p_{2}$ and $p_{3}$, respectively.
All the external momentum are on their mass-shell
and in order to simplify the evaluation the mass of the strange quark is neglected
\begin{equation}
p_{1}^{2}=m_{H}^{2},\quad p_{2}^{2}
=m_{b}^{2},\quad p_{3}^{2}=m_{s}^{2}=0.   \label{threeonshell}
\end{equation}
The equations of the motion for the bottom and the strange quarks are also needed
\begin{equation}
\bar{u}(p_{2})\slashed{p}_{2}=m_{b}\bar{u}(p_{2}),\quad
\slashed{p}_{3}v(p_{3})=0.  \label{equatiosofmotion}
\end{equation}
The left- and right-handed projection matrices $P_{L}$ and $P_{R}$
in Eq.~(\ref{gourponediagrama}) are
\begin{eqnarray}
P_{L}=\frac{1}{2}(1-\gamma_{5}),\quad \quad
P_{R}=\frac{1}{2}(1+\gamma_{5}). \label{leftandrighthanded}
\end{eqnarray}
Since
\begin{equation}
P_{L}\gamma_{\alpha}=\gamma_{\alpha}P_{R},\quad \quad
P_{R}\gamma_{\alpha}=\gamma_{\alpha}P_{L}, \label{cummutatingofleftandright}
\end{equation}
Eq.~(\ref{gourponediagrama}) is simplified to
\begin{equation}
M_{a}=-\frac{g_{2}^{2}(\sqrt{2}G_{f})^{1/2}}{m_{b}}
\sum_{q}V^{*}_{qb}V_{qs}\int\frac{\D^{4}k}{(2\pi)^{4}}
\frac{\bar{u}(p_{2})[-m_{b}\slashed{p}_{3}
+(\slashed{p}_{3}+m_{b})\slashed{k})]P_{L}v(p_{3})}{D_{2}D_{3}}. \label{simplifiedamplitudea}
\end{equation}
Before proceeding with the subsequent evaluation,
it is necessary to illustrate that the orthogonality
relation of the CKM matrix~\cite{Branco:1999fs}
\begin{equation}
V_{us}V^{*}_{ub}+V_{cs}V^{*}_{cb}+V_{ts}V^{*}_{tb}=0.  \label{CKMuorthogonality relation}
\end{equation}
does not appear in Eq.~(\ref{simplifiedamplitudea}),
otherwise the whole amplitude will vanish.
The reason is that since all the quarks in the propagators
in Eq.~(\ref{simplifiedamplitudea}) are massive,
the integrals are different from each other,
so that the coefficient of the product $V_{qs}V^{*}_{qb}(q=u, c, t)$
is not a common factor, and hence we get a nonzero contribution.
An analogous discussion can be applied to diagrams (c) and (d) in the second row
of Fig.~\ref{twotypediagofthestandardmodel}.

Performing the integral in \eqref{simplifiedamplitudea}
using dimensional regularization~\cite{tHooft:1972tcz, tHooft:1973mfk},
we obtain $M_{a}$ in the modified minimal
subtraction~($\overline{{\rm{MS}}}$) scheme
\begin{eqnarray}
M_{a}&=&4(\sqrt{2}G_{f})^{3/2}m_{W}^{2}
\sum_{q}V^{*}_{qb}V_{qs}\Big\{[B_{0}(m_{q})-B_{1}(m_{q})]
\bar{u}(p_{2})\slashed{p}_{1}P_{L}v(p_{3})\nonumber\\
&+&[B_{1}(m_{q})-B_{0}(m_{q})]
\bar{u}(p_{2})\slashed{p}_{2}P_{L}v(p_{3})\Big\},  \label{finalformofMa}
\end{eqnarray}
where $B_{0}$ and $B_{1}$ are the Passarino-Veltman
functions~\cite{tHooft:1978jhc, Passarino:1978jh} defined in Appendix~\ref{PVcoefficients}.
Imposing momentum conservation $p_{1}=p_{2}+p_{3}$
and Eq.~(\ref{equatiosofmotion}), it is evident that $M_{a}=0$.
A similar analysis can be applied to the evaluation of diagram (b)
and we also find $M_{b}=0$. Thus we do not need the
numerical value of $B_{0}$ and $B_{1}$,
but for completeness we provide the analytic expressions in  Appendix~\ref{PVcoefficients}.

The amplitude corresponding to diagram (c)
of Fig.~\ref{twotypediagofthestandardmodel} is
\begin{eqnarray}
M_{c}&=&-\frac{g_{2}^{2}(\sqrt{2}G_{f})^{1/2}}{2}
\sum_{q}m_{q}V^{*}_{qb}V_{qs}\int\frac{\D^{4}k}{(2\pi)^{4}}
\frac{1}{D_{1}D_{2}D_{3}}\nonumber\\
&\times&\bar{u}(p_{2})\gamma_{\mu}P_{L}(\slashed{k}+m_{q})
(\slashed{p}_{1}-\slashed{k}+m_{q})
\gamma^{\mu}P_{L}v(p_{3}), \label{firsttypeoftheSM}
\end{eqnarray}
where the three denominators are given by
\begin{eqnarray}
D_{1}&=&k^{2}-m_{q}^{2}+i\varepsilon,\nonumber\\
D_{2}&=&(p_{1}-k)^{2}-m_{q}^{2}+i\varepsilon,\nonumber\\
D_{3}&=&(p_{2}-k)^{2}-m_{W}^{2}+i\varepsilon.   \label{denominatorofdiagcfigone}
\end{eqnarray}
Carrying out the integral in Eq.~(\ref{firsttypeoftheSM}),
we can express Eq.~(\ref{firsttypeoftheSM}) through
the Passarino-Veltman function $C_{0}$
\begin{equation}
M_{c}=4m_{b}m_{W}^{2}(\sqrt{2}G_{f})^{3/2}
\Big[\sum_{q}m^{2}_{q}V^{*}_{qb}V_{qs}\,
C_{0}(m_{q},m_{q},m_{W})\Big]\bar{u}(p_{2})
P_{L}v(p_{3}).  \label{momentumintedamplitudema}
\end{equation}
It is straightforward to write down the amplitude of
diagram (d) in Fig.~\ref{twotypediagofthestandardmodel}
\begin{eqnarray}
M_{d}&=&-2g_{2}^{2}m_{W}^{2}(\sqrt{2}G_{f})^{1/2}
\sum_{q}V^{*}_{qb}V_{qs}
\int \frac{\D^{4}k}{(2\pi)^{4}}
\frac{\bar{u}(p_{2})(\slashed{k}-\slashed{p}_{2})
P_{L}v(p_{3})}{D_{1}D_{2}D_{3}},  \label{orignalexpreofmb}
\end{eqnarray}
where the three denominators are
\begin{eqnarray}
D_{1}&=&k^{2}-m_{W}^{2}+i\varepsilon,\nonumber\\
D_{2}&=&(p_{1}-k)^{2}-m_{W}^{2}+i\varepsilon,\nonumber\\
D_{3}&=&(p_{2}-k)^{2}-m_{q}^{2}+i\varepsilon.      \label{thrdenomanatorsofamb}
\end{eqnarray}
Performing the integral over $k$ using dimensional regularization,
we obtain
\begin{eqnarray}
M_{d}
&=&-8m_{b}m^{4}_{W}(\sqrt{2}G_{f})^{3/2}
\sum_{q}V^{*}_{qb}V_{qs}
\Big[C_{1}(m_{W},m_{W},m_{q})
+C_{2}(m_{W},m_{W},m_{q})\nonumber\\
&-&C_{0}(m_{W},m_{W},m_{q})\Big]
\bar{u}(p_{2})P_{L}v(p_{3}),       \label{secondtypeoftheSM}
\end{eqnarray}
where the explicit expressions of $C_{1}$ and $C_{2}$ can be
found in Appendix~\ref{PVcoefficients}.
Combining Eqs.~(\ref{momentumintedamplitudema}) and (\ref{secondtypeoftheSM}),
and using Eq.~(\ref{equatiosofmotion}),
we obtain the following $H\rightarrow b\bar{s}$ decay amplitude at leading order in the SM
\begin{eqnarray}
M_{{\rm{SM}}}&=&M_{c}+M_{d}\nonumber\\
&=&4m_{b}m^{2}_{W}(\sqrt{2}G_{f})^{3/2}
(\mathcal{A}_{1}+\mathcal{A}_{2})\bar{u}(p_{2})P_{L}v(p_{3}),   \label{amplitudeofdiagramandb}
\end{eqnarray}
where the two dimensionless constants are defined as
\begin{eqnarray}
\mathcal{A}_{1}&=&\sum_{q}
V^{*}_{qb}V_{qs}
m_{q}^{2}\,C_{0}(m_{q},m_{q},m_{W}), \nonumber\\
\mathcal{A}_{2}&=&-2m_{W}^{2}\sum_{q}V^{*}_{qb}V_{qs}
\Big[C_{1}(m_{W},m_{W},m_{q})
+C_{2}(m_{W},m_{W},m_{q})-
C_{0}(m_{W},m_{W},m_{q})\Big].      \label{twocoefficientsAoneandAtwo}
\end{eqnarray}
Summing over the spins of the $b$ and $\bar{s}$ for Eq.~(\ref{amplitudeofdiagramandb}) yields
\begin{equation}
|M_{{\rm{SM}}}|^{2}=32\sqrt{2}m_{b}^{2}G_{f}^{3}m_{W}^{4}(m_{H}^{2}-m_{b}^{2})
|\mathcal{A}_{1}+\mathcal{A}_{2}|^{2}.  \label{finalstatesspinsummed}
\end{equation}
Then the decay width can be evaluated through the following expression
\begin{equation}
\Gamma(H\rightarrow b\bar{s})
=\frac{N_{C}(m_{H}^{2}-m_{b}^{2})}{8\pi m_{H}^{3}}|M_{{\rm{SM}}}|^{2},   \label{decayrateinSM}
\end{equation}
where $N_{C}$ is the number of quark colors, and the incoherent
sum of the two final states $H\rightarrow b\bar{s}$ and $H\rightarrow \bar{b}s$
is considered. Replacing $V_{qs}$ by $V_{qd}$ in Eq.~(\ref{decayrateinSM}),
we also obtain the decay width for $H\rightarrow b\bar{d}$.
The other two flavor changing decays, $H\rightarrow s\bar{d}, c\bar{u}$,
will be not explored in this paper. The reason is that since we assume $m_{s}=0$,
the amplitude of $H\rightarrow s\bar{d}$ vanishes
which can be inferred from Eq.~(\ref{finalstatesspinsummed})
through replacing $m_{b}$ by $m_{s}$. Furthermore,
we do not explore the top partner effects on the $c\bar{u}$ final state
because it is a next-to-leading order effect
in the model used in this paper, and is hence
beyond the leading-order scope of this work.
Although the $c\bar{u}$ final state can be analyzed
at leading order in the context of models containing
down-type vector-like singlet quark, investigating models
of this type is also beyond the scope the present work.

\subsection{determination of the
quark mass parameters }   \label{massparameters}

There are four quark masses to be fixed the Eq.(\ref{decayrateinSM}), i.e.,
the masses of the $u$, $c$, $t$ and $b$.
Following the convention in Refs.~\cite{Aranda:2020tqw,
Kamenik:2024mpe, Dawson:2015oha},\footnote{In Refs.~\cite{Aranda:2020tqw, Kamenik:2024mpe},
the pole mass of the top quark is used for evaluating the
branching ratio of $H\rightarrow b\bar{s}$ at leading order in SM,
while in Ref.~\cite{Dawson:2015oha},
the pole mass of the top quark is applied to evaluate
the invariant mass distribution of $\D \sigma/\D m_{hh}$ in the
double Higgs production process $gg\rightarrow hh$.}
we use the pole mass for top quark in the numerical evaluation.
While for the other three quarks, the running mass
at the scale $m_{H}$ will be employed in the numerical evaluation.
The evolution of $\overline{m}_{Q}(\overline{m}_{Q})$ (as given in Ref.~\cite{ParticleDataGroup:2022pth})
upwards to some higher renormalization scale $\mu$ is determined by
\begin{equation}
\overline{m}_{Q}(\mu)=\overline{m}_{Q}(\overline{m}_{Q})
\frac{c\left[\dfrac{\alpha_{s}(\mu)}{\pi}\right]}
{c\left[\dfrac{\alpha_{s}(\overline{m}_{Q})}{\pi}\right]},   \label{runningatanymu}
\end{equation}
where the functions $c(x)$ are known up to
three loops~\cite{Chetyrkin:1997dh, Vermaseren:1997fq}\footnote{The evolution functions are first derived in
Refs.~\cite{Chetyrkin:1997dh, Vermaseren:1997fq},
here they are cited from Eq.~(11) of Ref.~\cite{Spira:1997dg}. }
\begin{eqnarray}
c(x)&=&\Big(\frac{9}{2}x\Big)^{4/9}\big(1+0.895x+1.371x^{2}+1.952x^{3}\big),\quad
\text{for} \quad m_{s}<\mu<m_{c} \nonumber\\
c(x)&=&\Big(\frac{25}{6}x\Big)^{12/25}\big(1+1.014x+1.389x^{2}+1.091x^{3}\big),\quad
\text{for} \quad m_{c}<\mu<m_{b} \nonumber\\
c(x)&=&\Big(\frac{23}{6}x\Big)^{12/23}\big(1+1.175x+1.501x^{2}+0.1725x^{3}\big),    \quad
\text{for} \quad m_{b}<\mu<m_{t} \nonumber\\
c(x)&=&\Big(\frac{7}{2}x\Big)^{4/7}\big(1+1.398x+1.793x^{2}-0.6834x^{3}\big).\quad
\text{for} \quad m_{t}<\mu      \label{runingfunctionc}
\end{eqnarray}
The analytic expression for $\alpha_{s}(\mu)$ is presented in Appendix A.

When we apply Eq.~(\ref{runningatanymu}) and the piecewise
evolution functions in Eq.~(\ref{runingfunctionc}) to
evaluate $\overline{m}_{Q}(m_{H})$,
we must cope with the threshold effects~\cite{Steele:1998qls}.
For instance, if we run $\overline{m}_{c}(\overline{m}_{c})$ to
$\overline{m}_{c}(m_{H})$, we should carefully deal with the effects
when the scale passes through $\overline{m}_{b}(\overline{m}_{b})$.
A way out of this dilemma is as follows. Since there is no threshold
between $\mu=\overline{m}_{b}(\overline{m}_{b})$ and $\mu=m_{H}$,
we can directly obtain $\overline{m}_{b}(m_{H})$
from $\mu=\overline{m}_{b}(\overline{m}_{b})$
via the third function in Eq.~(\ref{runingfunctionc}).
Then combined with the scale-independent ratio~\cite{ParticleDataGroup:2022pth}
\begin{equation}
\frac{\overline{m}_{b}}{\overline{m}_{c}}=4.584\pm 0.007.   \label{ratioofrunningmasses}
\end{equation}
we can obtain $\overline{m}_{c}(m_{H})$. The results
are tabulated in Table~\ref{differentmassofcandbquarks}.

Similarly, to avoid threshold-matching issues for
the $u$ quark mass (e.g., Ref.~\cite{ParticleDataGroup:2022pth}
provides the $u$ mass at a scale of $2\,{\rm GeV}$),
we again use scale-independent mass ratios.
Defining
\begin{equation}
\overline{m}_{n}=\frac{\overline{m}_{u}+\overline{m}_{d}}{2}, \label{definingmn}
\end{equation}
and by making use of the scale-independent ratios~\cite{ParticleDataGroup:2022pth}
\begin{equation}
\xi_{cs}=\frac{\overline{m}_{c}}{\overline{m}_{s}}=11.76^{+0.05}_{-0.10},\quad
\xi_{ud}=\frac{\overline{m}_{u}}{\overline{m}_{d}}=0.474^{+0.056}_{-0.074},\quad
\xi_{sn}=\frac{\overline{m}_{n}}{\overline{m}_{u}}=27.33^{+0.67}_{-0.077}, \label{scaleindependentratios}
\end{equation}
we get
\begin{equation}
\frac{\overline{m}_{n}}{\overline{m}_{u}}=\frac{2\xi_{ud}}{1+\xi_{ud}}.  \label{ratioofnandu}
\end{equation}
Rearranging the ratio $\overline{m}_{u}/\overline{m}_{c}$ into the following form
\begin{equation}
\frac{\overline{m}_{u}}{\overline{m}_{c}}
=\frac{\overline{m}_{u}}{\overline{m}_{n}}
\frac{\overline{m}_{n}}{\overline{m}_{s}}
\frac{\overline{m}_{s}}{\overline{m}_{c}}=
\frac{2\xi_{ud}}{1+\xi_{ud}}
\frac{1}{\xi_{sn}}\frac{1}{\xi_{cs}}=0.0020, \label{ratioofmuandmc}
\end{equation}
then combining with the value $\overline{m}_{c}(m_{H})$ in Table.~\ref{differentmassofcandbquarks},
we find the running mass of up quark at scale $\mu=m_{H}$
\begin{equation}
\overline{m}_{u}(m_H)=1.22\pm 0.01\, {\rm MeV}.  \label{runningmassofup}
\end{equation}

\begin{table}
\begin{center}
\caption{The pole masses of $c$ and $b$ as well as
their running masses at the scale $\overline{m}_{Q}(\overline{m}_{Q})$
and at the scale $m_{H}=125.25\,{\rm{GeV}}$
for $\Lambda_{{\rm{QCD}}}=0.208\,{\rm{GeV}}$.
Ref.~\cite{ParticleDataGroup:2022pth}
values are used for the pole masses and $\overline{m}_{Q}(\overline{m}_{Q})$.
The unit of the masses is ${\rm{GeV}}$. }
\label{differentmassofcandbquarks}
\begin{tabular}{cccc}
\hline
$Q$\quad\quad\quad\quad\quad & \quad\quad\quad\quad\quad$m_{Q}$
&\quad\quad\quad\quad\quad$\overline{m}_{Q}(\overline{m}_{Q})$
&\quad\quad\quad\quad\quad $\overline{m}_{Q}(m_{H})$\\
\hline
$c$\quad\quad\quad\quad\quad &$\quad\quad\quad\quad\quad1.67\pm0.07$
&\quad\quad\quad\quad\quad$1.27\pm 0.02$
&\quad\quad\quad\quad\quad$0.609^{+0.006}_{-0.003}$\\
$b$\quad\quad\quad\quad\quad&\quad\quad\quad\quad\quad$4.78\pm0.06$
&\quad\quad\quad\quad\quad$4.18^{+0.03}_{-0.02}$
&\quad\quad\quad\quad\quad $2.793^{+0.014}_{-0.016}$\\
\hline
\end{tabular}
\end{center}
\end{table}

\subsection{numerical results and discussion}
For definiteness, we list all the masses needed in the evaluation
\begin{eqnarray}
&&m_{u}\left(m_H\right)=1.22\pm 0.01\,{\rm{MeV}},
\quad m_{c}\left(m_H\right)=0.609^{+0.006}_{-0.003}{\rm GeV},\quad
m_{b}\left(m_H\right)=2.793^{+0.014}_{-0.016}{\rm GeV}\nonumber\\
&&m_{t}=172.69\pm0.30\,{\rm{GeV}},\quad
m_{W}=80.377\pm0.012\,{\rm{GeV}},\nonumber\\
&& m_{H}=125.25\pm0.017\,{\rm{GeV}},\label{massesofquarksandbosonsnew}  \end{eqnarray}

As noted earlier, for the top quark we employ the pole mass,
following the conventions of Ref.~\cite{Aranda:2020tqw,Kamenik:2024mpe,Dawson:2015oha}
in the evaluation the $H \rightarrow b\bar{s}$
at leading order in SM and the invariant mass distribution in $gg \rightarrow hh$ process.

The CKM elements are expressed by the Wolfenstein paramaterization~\cite{Wolfenstein:1983yz}
\begin{gather}
V_{ud}=1-\frac{1}{2}\lambda^{2}-\frac{1}{8}\lambda^{4}+\mathcal{O}(\lambda^{6}), \nonumber\\
V_{us}=\lambda+\mathcal{O}(\lambda^{7}),\nonumber\\
V_{ub}=A\lambda^{3}(\rho-i\eta),\nonumber\\
V_{cd}=-\lambda+\frac{1}{2}A^{2}\lambda^{2}[1-2(\rho+i\eta)]+\mathcal{O}(\lambda^{7}), \nonumber\\
V_{cs}=1-\frac{\lambda^{2}}{2}+\frac{\lambda^{4}}{8}(1+4A^{2})+\mathcal{O}(\lambda^{6}),\nonumber\\
V_{cb}=A\lambda^{2}+\mathcal{O}(\lambda^{8}),\nonumber\\
V_{td}=A\lambda^{3}[1-(\rho+i\eta)(1-\frac{1}{2}\lambda^{2})]+\mathcal{O}(\lambda^{7}),\nonumber\\
V_{ts}=-A\lambda^{2}+\frac{1}{2}(1-2\rho)\lambda^{4}-i\eta A \lambda^{4}
+\mathcal{O}(\lambda^{6}),\nonumber\\
V_{tb}=1-\frac{1}{2}A^{2}\lambda^{4}+\mathcal{O}(\lambda^{6}),  \label{relatedCKMelements}
\end{gather}
with
\begin{equation}
\bar{\rho}=\rho(1-\frac{\lambda^{2}}{2}),\quad\quad
\bar{\eta}=\eta(1-\frac{\lambda^{2}}{2}),  \label{definitionofrhobaretabar}
\end{equation}
The up-to-date fit of the above parameters are~\cite{ParticleDataGroup:2022pth}
\begin{eqnarray}
&&\lambda=0.22500\pm0.00067,
\quad\quad  A=0.826^{+0.018}_{-0.015},\nonumber\\
&&\bar{\rho}=0.159\pm0.010,
\quad\quad \bar{\eta}=0.348\pm0.010.\label{uptodatefitCKMpara}
\end{eqnarray}

\begin{table}
\begin{center}
\caption{Numerical results for $\mathcal{F}^{{\rm SM}}_{i}~(i=1,2)$
in the SM using central values for all parameters.}
\label{comparsionofcoefficientsofsm}
\begin{tabular}{ccc}
\hline
Decay &$\mathcal{F}^{{\rm{SM}}}_{1}$ & $\mathcal{F}^{{\rm{SM}}}_{2}$\\
\hline
$H\rightarrow b\bar{s}$\quad\quad\quad&$-2.43\times10^{-6}-4.46\times10^{-8}i$\quad\quad\quad
&$-4.00\times10^{-6}-1.61\times10^{-7}i$\\
$H\rightarrow b\bar{d}$\quad\quad\quad&$4.68\times10^{-7}-1.22\times10^{-7}i$\quad\quad\quad
&$5.87\times10^{-7}-6.61\times10^{-7}i$\\
\hline
\end{tabular}
\end{center}
\end{table}

Taking the central values of the parameters in the
expressions for $\mathcal{A}_{1}$ and $\mathcal{A}_{2}$,
we can compare their contribution to the total amplitude of the SM. Defining
\begin{equation}
\mathcal{F}^{{\rm{SM}}}_{i}=4m_{b}m^{2}_{W}(\sqrt{2}G_{f})^{3/2}
\mathcal{A}_{i},\quad \quad \quad i=\{1,\,2\},  \label{relativesizeoftheSMdiagrams}
\end{equation}
leads to the results in Table~\ref{comparsionofcoefficientsofsm}.
The values of $\mathcal{F}^{{\rm{SM}}}_{1}$ and
$\mathcal{F}^{{\rm{SM}}}_{2}$ indicate that the main contribution
for both decays are from diagram (d) which can be understood becausethe $g_{Hff}$ coupling
is less than the $g_{HWW}$ coupling. Substituting all the parameters into Eq.~(\ref{decayrateinSM}),
we obtain the following SM Higgs decay width at the scale $\mu=m_{H}$
for the two decays\footnote{The numerical evaluation is implemented via Mathematica:
https://www.wolfram.com/mathematica/.}
\begin{equation}
\Gamma(H\rightarrow b\bar{s})=6.17
\times10^{-7}\,{\rm{MeV}},\quad
 \Gamma(H\rightarrow b\bar{d})=2.58
\times10^{-8}\,{\rm{MeV}}.    \label{decaywidthinstandardmodel}
\end{equation}
Combining Eq.(\ref{decaywidthinstandardmodel}) with the total width of the Higgs boson
$\Gamma_{H}=3.2^{+2.4}_{-1.7}{\rm{MeV}}$~\cite{ParticleDataGroup:2022pth},
we obtain the branching fractions
\begin{equation}
\frac{\Gamma(H\rightarrow b\bar{s})}{\Gamma_{H}}
=1.93\times10^{-7},\quad
\frac{\Gamma(H\rightarrow b\bar{d})}{\Gamma_{H}}
=8.05\times10^{-9},    \label{fractioninstandardmodel}
\end{equation}
in good agreement with the results based
on the one-loop SM evaluation in
Refs.~\cite{Benitez-Guzman:2015ana,
Aranda:2020tqw, Farrera:2020bon, Arco:2023hmz}.\footnote{The
branching ratios in Ref.~\cite{Farrera:2020bon} are evaluated
in an universal extra dimension model.
Unlike the four-dimensional SM and its extensions,
there are more Feynman diagrams that contribute to
the amplitudes of $H\rightarrow b\bar{s}, b\bar{d}$
decays due to the Kaluza-Klein (KK) excited-mode quark fields.}
The branching ratios of the two processes are lower
than all the observed channels of the Higgs decay on the LHC,
thus it is challenging to detect them.

We now consider the process $H\rightarrow b\bar{s}$
as specific example to explore possible ways to enhance amplitude.
One obvious choice is the next-to-leading
order corrections induced by QCD. But at the scale of $m_{H}$
where the strong coupling constant $\alpha_{s}\sim 0.1$,
the squared amplitude will be suppressed compared to leading order.
Thus we do not take it as a viable way to
enhance the results in Eq.~(\ref{decaywidthinstandardmodel}).
Another possibility is exploring
contributions from new particles.
As mentioned in the introduction, the vector-like singlet
fermion model is promising because the top partner also
can contribute to the decay $H\rightarrow b\bar{s}$ at leading order.
This implies that we can view this decay mode as a
sensitive probe to explore the effects of the
vector-like singlet top partner.
This will be presented in the next section.

\begin{figure}
\begin{center}
\includegraphics[scale=0.50]{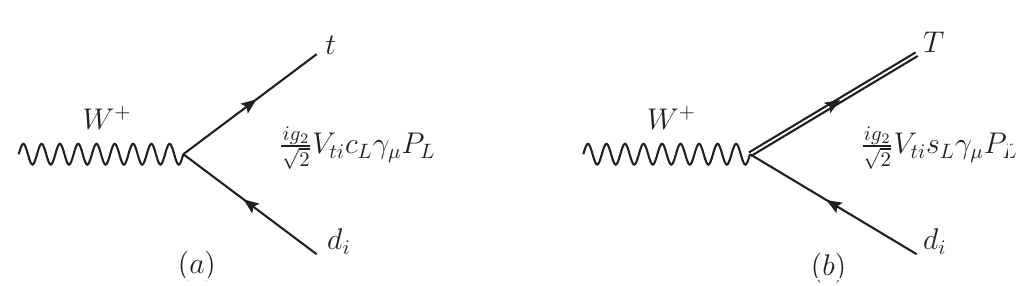}
\caption{Two types of couplings to $W$ induced by the top partner.}\label{twotypevertexofwboson}
\end{center}
\end{figure}

\begin{figure}
\begin{center}
\includegraphics[scale=0.50]{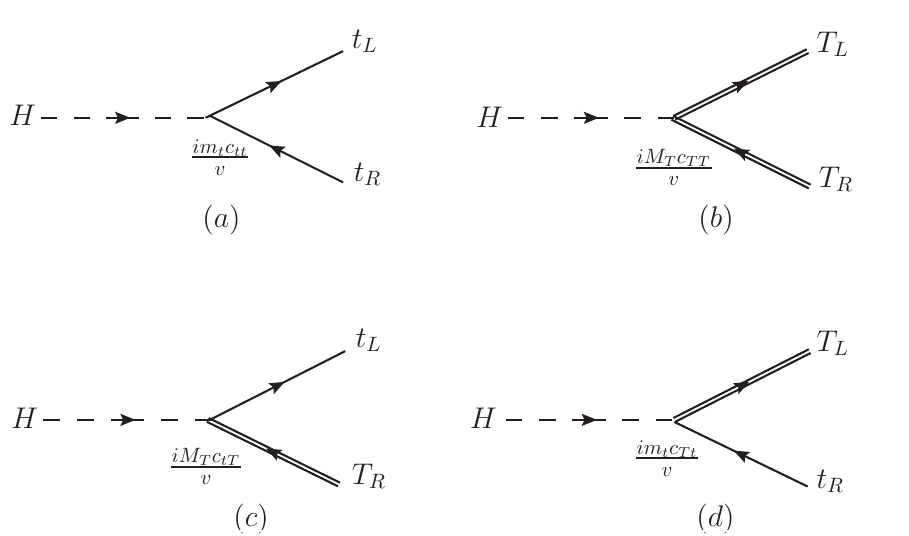}
\caption{Couplings of the top and its vector-like singlet
partner to the Higgs boson.}  \label{fourtypevertexofTt}
\end{center}
\end{figure}

\section{evaluating the branching ratios of
the processes $H\rightarrow b\bar{s},\,b\bar{d}$
with inclusion of the vector-like singlet top partner}
\label{BR_section}

\subsection{the amplitude formulae}

\begin{figure}
\begin{center}
\includegraphics[scale=0.50]{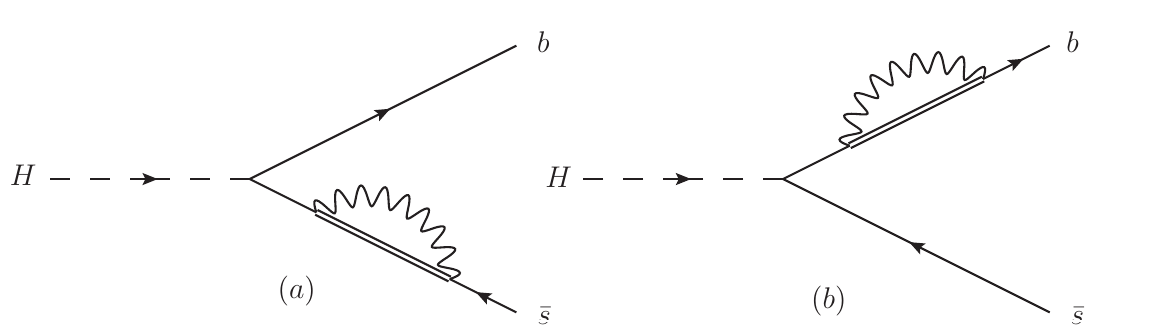}
\caption{The negligible one-loop diagrams with
inclusion of the top partner.}\label{neglectedtoppartner}
\end{center}
\end{figure}

\begin{figure}
\begin{center}
\includegraphics[scale=0.50]{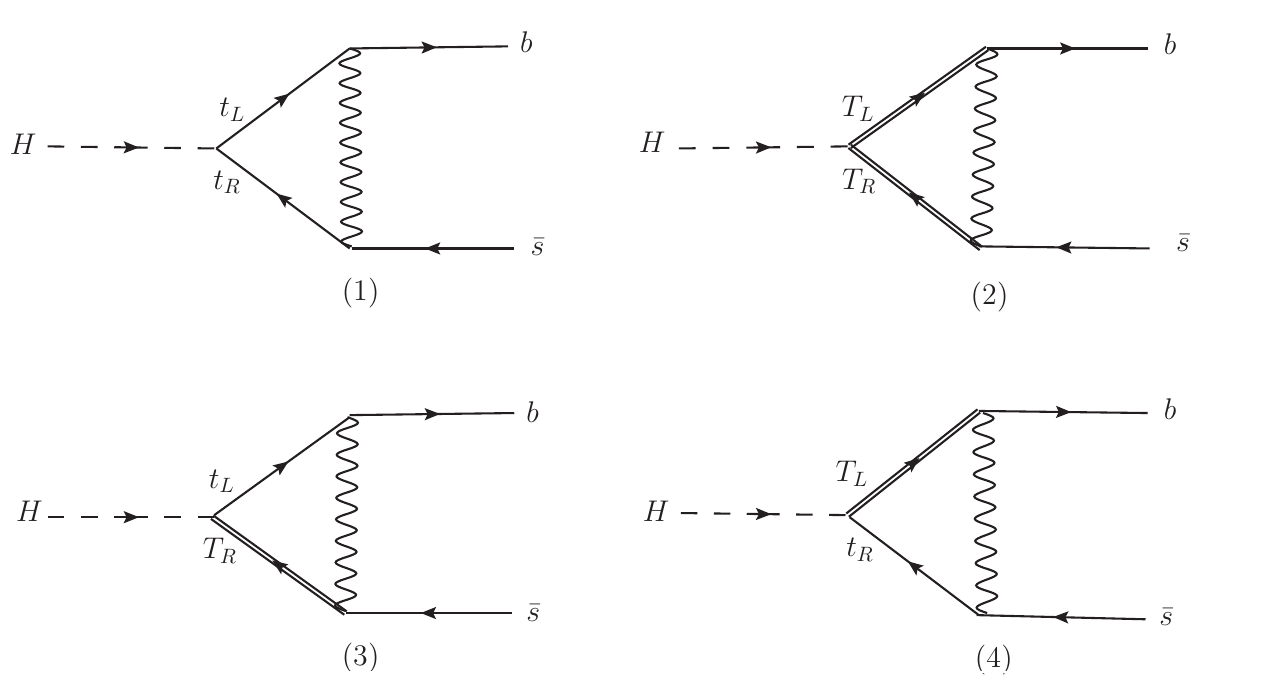}
\includegraphics[scale=0.50]{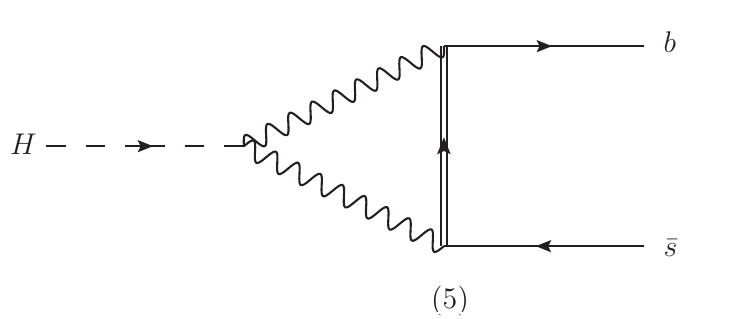}
\caption{One-loop diagrams contributing to $H\rightarrow b \bar{s}$
with inclusion of the top partner.}  \label{toppartnercontributon}
\end{center}
\end{figure}

The Lagrangian density describing the interaction of top partner with
the $W^{\pm}$ is~\cite{Kamenik:2017tnu}
\begin{equation}
\mathcal{L}_{W}=-\frac{g_{2}}{\sqrt{2}}
V_{ti}(c_{L}\bar{t}\,\slashed{W}^{+}P_{L}d_{i}
+s_{L}\overline{T}\slashed{W}^{+}P_{L}d_{i}), \label{lagrangianofwbosonandTt}
+\rm{h.c.}
\end{equation}
where $V_{ti}~(i=d,s,b)$ are the elements of the CKM matrix,
and $\{c_{L},s_{L}\}$ are  respectively abbreviations for $\{\cos\theta_{L},\sin\theta_{L}\}$.
The mixing between the top quark and its singlet partner is~\cite{Dawson:2012di, Fox:2018ldq}
\begin{equation}
\begin{pmatrix}
t_{L}\\
T_{L}
\end{pmatrix}
=\begin{pmatrix}
\cos\theta_{L}&-\sin\theta_{L}\\
\sin\theta_{L}&\cos\theta_{L}
\end{pmatrix}
\begin{pmatrix}
\mathcal{T}^{1}_{L}\\
\mathcal T^{2}_{L}
\end{pmatrix},\quad\quad
\begin{pmatrix}
t_{R}\\
T_{R}
\end{pmatrix}
=\begin{pmatrix}
\cos\theta_{R}&-\sin\theta_{R}\\
\sin\theta_{R}&\cos\theta_{R}
\end{pmatrix}
\begin{pmatrix}
\mathcal{T}^{1}_{R}\\
\mathcal T^{2}_{R}
\end{pmatrix},  \label{topandthepartnermixing}
\end{equation}
where $t_{L, R}$ and $T_{L,R}$ are the respective mass eigenstates
of top quark and the top partner, while
$\mathcal{T}^{1}_{R, L}$ and $\mathcal{T}^{2}_{R, L}$
are the corresponding weak eigenstates.
The interactions with the SM Higgs boson are~\cite{Dawson:2012di}
\begin{equation}
\mathcal{L}=-\frac{m_{t}}{v}c_{tt}\overline{t_{L}}t_{R}h
-\frac{M_{T}}{v}c_{TT}\overline{T_{L}}T_{R}h
-\frac{M_{T}}{v}c_{tT}\overline{t_{L}}T_{R}h
-\frac{m_{t}}{v}c_{Tt}\overline{T_{L}}t_{R}h+\rm{h.c.}, \label{lagrangianofhiggsandTt}
\end{equation}
with
\begin{equation}
c_{tt}=c_{L}^{2},\quad
c_{TT}=s_{L}^{2},\quad c_{tT}=c_{Tt}=s_{L}c_{L},  \label{definitionofclsl}
\end{equation}
From Eq.~(\ref{lagrangianofwbosonandTt}) and Eq.~(\ref{lagrangianofhiggsandTt}),
we can obtain six new types of vertices
which are depicted in Fig.~\ref{twotypevertexofwboson}
and in Fig.~\ref{fourtypevertexofTt}. At leading order,
the diagrams contributing to the amplitude
of $H\rightarrow b\bar{s}$ are depicted
in Fig.~\ref{neglectedtoppartner} and in Fig.~\ref{toppartnercontributon}.
By using the equation of motion of the $s$ quark, we conclude
that the contributions from the diagrams in Fig.~\ref{neglectedtoppartner}
vanish, so the top partner contributes to the
amplitude through the diagrams in Fig.~\ref{toppartnercontributon}.

The amplitude of the first diagram in Fig.~\ref{toppartnercontributon} is
\begin{eqnarray}
M_{1}&=&\frac{g^{2}_{2}}{2}
\frac{m_{t}}{v}V^{*}_{tb}V_{ts}c_{tt}c_{L}^{2}
\int\frac{\D^{4}k}{(2\pi)^{4}}
\frac{\bar{u}(p_{2})
\gamma_{\mu}\gamma_{L}(\slashed{k}+m_{t})
P_{R}(\slashed{p}_{1}
-\slashed{k}+m_{t})\gamma^{\mu}P_{L}v(p_{3})}
{D_{1}D_{2}D_{3}},  \label{firstamplituedincldingtoppartner}
\end{eqnarray}
where the denominators are
\begin{eqnarray}
D_{1}&=&k^{2}-m_{t}^{2}+i\varepsilon,\nonumber\\
D_{2}&=&(p_{1}-k)^{2}-m_{t}^{2}+i\varepsilon,\nonumber\\
D_{3}&=&(p_{2}-k)^{2}-m_{W}^{2}+i\varepsilon.    \label{threeprogantorsinmone}
\end{eqnarray}
By employing Eq.~(\ref{cummutatingofleftandright}), we obtain
\begin{eqnarray}
M_{1}&=&g_{2}^{2}\frac{m_{t}^{2}}{v}V^{*}_{tb}V_{ts}c_{tt}c_{L}^{2}
\int\frac{\D^{4}k}{(2\pi)^{4}}
\frac{\bar{u}(p_{2})\slashed{k}P_{L}v(p_{3})}
{D_{1}D_{2}D_{3}}.\label{originalmonebytoppartner}
\end{eqnarray}
After integrating over the momentum using dimensional regularization,
we obtain the following result
\begin{equation}
M_{1}=4Z_{1}m_{b}m_{W}^{2}(\sqrt{2}G_{f})^{3/2}
\big[C_{1}^{(1)}(m_{t},m_{t},m_{W})
+C_{2}^{(1)}(m_{t},m_{t},m_{W})\big]
\bar{u}(p_{2})P_{L}v(p_{3}),  \label{moneinducedbytoppartner}
\end{equation}
where the constant $Z_{1}$ is
\begin{equation}
Z_{1}=-m^{2}_{t}V^{*}_{tb}V_{ts}c_{tt}c_{L}^{2}, \label{zoneincldingtoppartner}
\end{equation}
 we have expressed the vacuum value of
 the Higgs boson in terms of the Fermi coupling constant,
 and $C_{1}^{(1)}$, $C_{1}^{(2)}$ are the Passarino-Veltman functions
 listed in Appendix~\ref{PVcoefficients}.
 The amplitudes of the next three diagrams can be obtained in a similar manner
\begin{eqnarray}
M_{2}&=&4Z_{2}m_{b}m_{W}^{2}(\sqrt{2}G_{f})^{3/2}
\big[C_{1}^{(2)}(M_{T},M_{T},m_{W})
+C_{2}^{(2)}(M_{T},M_{T},m_{W})\big]
\bar{u}(p_{2})P_{L}v(p_{3}),\nonumber\\
M_{3}&=&4Z_{3}m_{b}m_{W}^{2}(\sqrt{2}G_{f})^{3/2}
\big[C_{1}^{(3)}(m_{t},M_{T},m_{W})
+C_{2}^{(3)}(m_{t},M_{T},m_{W})\big]
\bar{u}(p_{2})P_{L}v(p_{3}),\nonumber\\
M_{4}&=&4Z_{4}m_{b}m_{W}^{2}(\sqrt{2}G_{f})^{3/2}
\big[C_{1}^{(4)}(m_{t},M_{T},m_{W})
+C_{2}^{(4)}(m_{t},M_{T},m_{W})\big]
\bar{u}(p_{2})P_{L}v(p_{3}), \label{toppartnermtwotofour}
\end{eqnarray}
with $Z_{i}~(i=2,3,4)$ defined as
\begin{equation}
Z_{2}
=-M_{T}^{2}V^{*}_{tb}V_{ts}s^{2}_{L}c_{TT},\quad
Z_{3}
=-M_{T}^{2}V^{*}_{tb}V_{ts}c_{L}s_{L}c_{tT},\quad
Z_{4}
=-m_{t}^{2}V^{*}_{tb}V_{ts}c_{L}s_{L}c_{Tt}.   \label{zotwotozfour}
\end{equation}
The amplitude of the last diagram in Fig.~\ref{toppartnercontributon}
is
\begin{eqnarray}
M_{5}&=&-2g_{2}^{2}m_{W}^{2}\sin^{2}\theta_{L}(\sqrt{2}G_{f})^{1/2}
V^{*}_{tb}V_{ts}
\int \frac{\D^{4}k}{(2\pi)^{4}}
\frac{\bar{u}(p_{2})(\slashed{k}-\slashed{p}_{2})
P_{L}v(p_{3})}{D_{1}D_{2}D_{3}},  \label{amoffifthtoppartner}
\end{eqnarray}
where the three denominators are
\begin{eqnarray}
D_{1}&=&k^{2}-m_{W}^{2}+i\varepsilon,\nonumber\\
D_{2}&=&(p_{1}-k)^{2}-m_{W}^{2}+i\varepsilon,\nonumber\\
D_{3}&=&(p_{2}-k)^{2}-M_{T}^{2}+i\varepsilon,      \label{denominatorofthefifth}
\end{eqnarray}
Completing the integral over $k$ using dimensional regularization, we obtain
\begin{equation}
M_{5}=4Z_{5}m_{b}m^{2}_{W}(\sqrt{2}G_{f})^{3/2}
\bar{u}(p_{2})P_{L}v(p_{3}),  \label{amplitudemfive}
\end{equation}
where $Z_{5} $ is given by
\begin{eqnarray}
Z_{5}=-2m^{2}_{W}\sin^{2}\theta_{L}
V^{*}_{tb}V_{ts}
\Big[C^{(5)}_{1}(m_{W},m_{W},M_{T})
&+&C^{(5)}_{2}(m_{W},m_{W},M_{T})
\nonumber\\
&-&
C^{(5)}_{0}(m_{W},m_{W},M_{T})\Big].
\label{momentumintedfifthtoppartner}
\end{eqnarray}

We can now form the total contributions
from the top partner to the amplitude of the process $H\rightarrow b\bar{s}$
\begin{equation}
M_{{\rm{VL}}}=\sum_{i=1}^{5}M_{i}
=4m_{b}m_{W}^{2}(\sqrt{2}G_{f})^{3/2}
\Big(\sum_{i=1}^{5}\mathcal{C}_{i}\Big)\bar{u}(p_{2})P_{L}v(p_{3}),  \label{totalamplitudebytoppartner}
\end{equation}
where the coefficients $\mathcal{C}_{i}$ are given by
\begin{eqnarray}
\mathcal{C}_{i}&=&Z_{i}\big[C_{1}^{(i)}
+C_{2}^{(i)}\big],\quad i=1,\,2,\,3,\,4\nonumber\\
\mathcal{C}_{5}&=&Z_{5}\Big[C^{(5)}_{1}(m_{W},m_{W},M_{T})
+C^{(5)}_{2}(m_{W},m_{W},M_{T})
-C^{(5)}_{0}(m_{W},m_{W},M_{T})\Big], \label{fourcoefficientsoftoppartner}
\end{eqnarray}
and for brevity the mass dependence in $C_{1}^{(i)}$ and
$C_{2}^{(i)}$ has been suppressed. Combining this result with Eq.~(\ref{amplitudeofdiagramandb})
yields the total amplitude
\begin{eqnarray}
M_{{\rm{tot}}}&=&M_{{\rm{SM}}}+M_{{\rm{VL}}}\nonumber\\
&=&4m_{b}m^{2}_{W}(\sqrt{2}G_{f})^{3/2}
\Big(\mathcal{A}_{1}+\mathcal{A}_{2}
+\sum_{i=1}^{5}\mathcal{C}_{i}\Big)\bar{u}(p_{2})P_{L}v(p_{3}).  \label{amplitudeincludingtopphobic}
\end{eqnarray}
Summing over the spins of the final quarks, we arrive at
\begin{eqnarray}
|M_{{\rm{tot}}}|^{2}&=&32\sqrt{2}m_{b}^{2}
G_{f}^{3}m_{W}^{4}(m_{H}^{2}-m_{b}^{2})
\Big|\mathcal{A}_{1}+\mathcal{A}_{2}
+\sum_{i=1}^{5}\mathcal{C}_{i}\Big|^{2}.  \label{totalamplitudes}
\end{eqnarray}
Substituting Eq.~(\ref{totalamplitudes}) into Eq.~(\ref{decayrateinSM}),
we can analyze the effects of the top partner
on the $H\rightarrow b\bar{s}$ branching ratio.
Finally, we note that the total amplitude of the
$H\rightarrow b\bar{d}$ can be obtained by replacing
$V_{ts}$ by $V_{td}$ in Eq.~(\ref{totalamplitudes}).

\subsection{numerical results and discussion}

\begin{table}
\begin{center}
\caption{Values of $\mathcal{F}^{{\rm{VL}}}_{i}~(i=1,\,2,...,5)$
for $H\rightarrow b\bar{s}$ decay for  $M_{T}=1500\,{\rm{GeV}}$ and
$\sin\theta_{L}=0.02$, with central values for all other parameters.}
\label{coefficientsofhtobswithvectortop}
\begin{tabular}{cc}
\hline
Coefficient & Value\\
\hline
$\mathcal{F}^{{\rm{VL}}}_{1}$&$-3.23\times10^{-8}-5.94\times10^{-10}i$\\
$\mathcal{F}^{{\rm{VL}}}_{2}$&$-1.19\times10^{-7}-2.19\times10^{-9}i$\\
$\mathcal{F}^{{\rm{VL}}}_{3}$&$-2.99\times10^{-4}-5.48\times10^{-6}i$\\
$\mathcal{F}^{{\rm{VL}}}_{4}$&$-3.95\times10^{-6}-7.26\times10^{-8}i$\\
$\mathcal{F}^{{\rm{VL}}}_{5}$&$-1.72\times10^{-6}-3.17\times10^{-8}i$\\
\hline
\end{tabular}
\end{center}
\end{table}

\begin{table}
\begin{center}
\caption{Values of $\mathcal{F}^{{\rm{VL}}}_{i}~(i=1,2,...,5)$
for the process $H\rightarrow b\bar{d}$ for $M_{T}=1500\,{\rm{GeV}}$
and $\sin\theta_{L}=0.02$, with central values for all other parameters.}
\label{fourcoeffofhtobdwithvectortop}
\begin{tabular}{cc}
\hline
Coefficient & Value\\
\hline
$\mathcal{F}^{{\rm{VL}}}_{1}$&$6.22\times10^{-9}-1.62\times10^{-9}i$\\
$\mathcal{F}^{{\rm{VL}}}_{2}$&$2.29\times10^{-8}-5.98\times10^{-9}i$\\
$\mathcal{F}^{{\rm{VL}}}_{3}$&$5.73\times10^{-5}-1.49\times10^{-5}i$\\
$\mathcal{F}^{{\rm{VL}}}_{4}$&$7.60\times10^{-7}-1.98\times10^{-7}i$\\
$\mathcal{F}^{{\rm{VL}}}_{5}$&$3.32\times10^{-7}-8.65\times10^{-8}i$\\
\hline
\end{tabular}
\end{center}
\end{table}

\begin{figure}
\begin{center}
\includegraphics[scale=0.75]{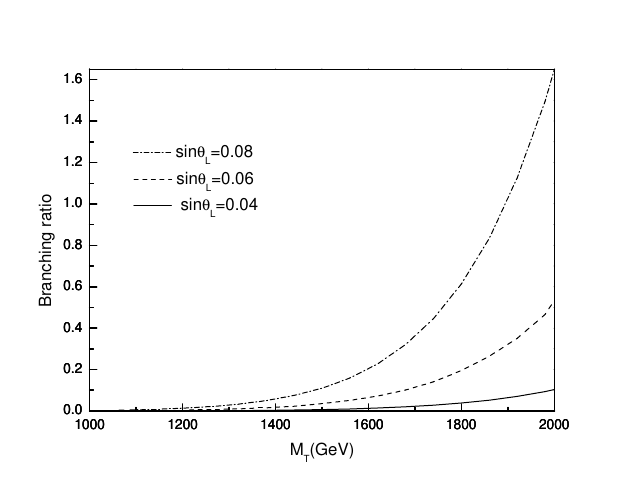}
\includegraphics[scale=0.75]{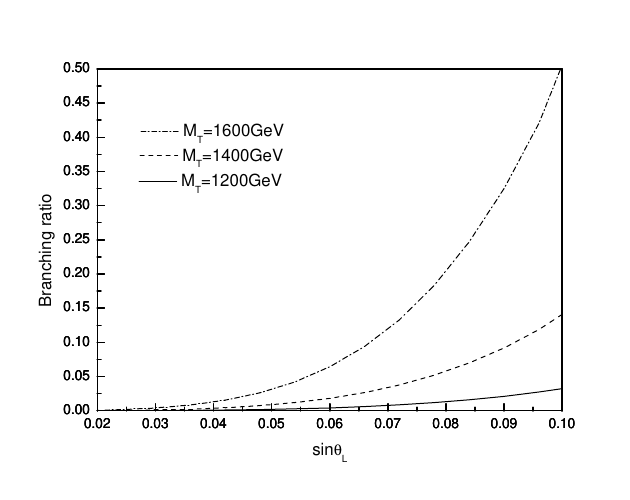}
\caption{The $H\rightarrow b\bar{s}$ branching ratios as
a function of the mass of the vector-like top partner
with selected $\sin\theta_{L}$ values (left)
and as a function of the mixing angle
with selected $M_{T}$ values (right).}  \label{varationoftheHtobsbarbranchfrac}
\end{center}
\end{figure}

\begin{figure}
\begin{center}
\includegraphics[scale=0.75]{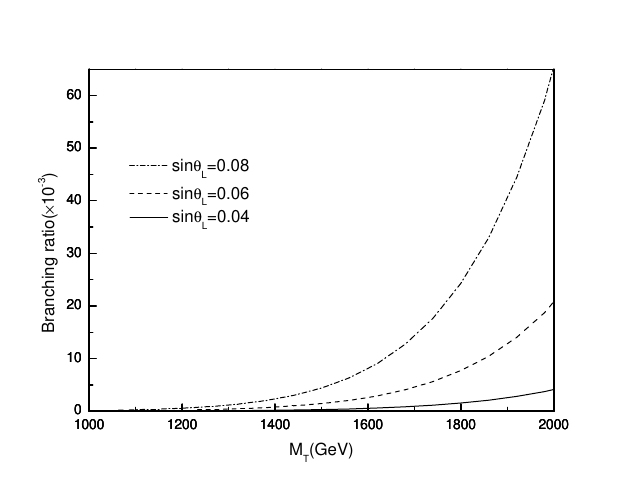}
\includegraphics[scale=0.75]{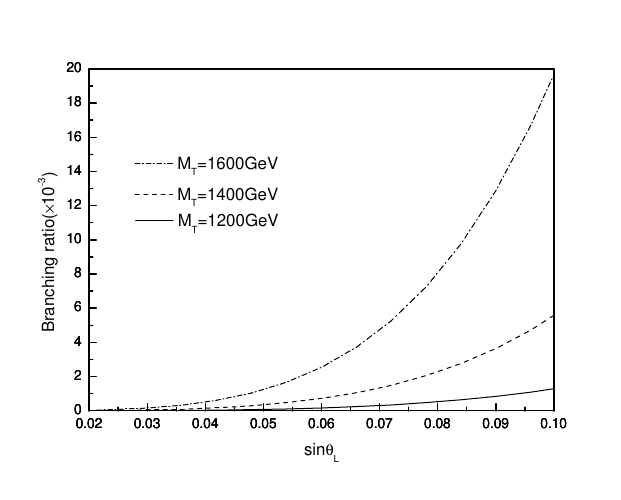}
\caption{The branching ratios of $H\rightarrow b\bar{d}$ as
a function of the mass of the vector-like
top partner with selected $\sin\theta_{L}$ values (left)
and as a function of the mixing angle
with selected $M_{T}$ values (right). }  \label{varationofHtobdbarbranchfrac}
\end{center}
\end{figure}

In order to the investigate mixing effects
on the branching ratios of the process $H\rightarrow b\bar{s}$
the branching ratios as a function of $M_{T}$ for selected values
$\sin\theta_{L}=\{0.04, 0.06, 0.08\}$ are shown
in the left panel of Fig.~\ref{varationoftheHtobsbarbranchfrac}.
Conversely, the branching ratios as a function of $\sin\theta_{L}$
for selected values $M_{T}=\{1200, 1400, 1600\,{\rm{GeV}\}}$
are presented in the right panel of
Fig.~\ref{varationoftheHtobsbarbranchfrac}.
Similarly, results for $H\rightarrow b\bar{d}$
are presented in Fig.~\ref{varationofHtobdbarbranchfrac}.
It is obvious that the branching ratios of both decays rise
quickly with $M_{T}$ or $\sin\theta_{L}$.
This behavior occurs because the couplings of the
top partner to the top quark and the Higgs boson
as well as to $W$ are proportional to the
product of $\sin\theta_{L}$ and $M_{T}$,
which can largely compensate for the suppression caused by $G_{f}$
and the CKM matrix. As a result, the amplitude grows rapidly,
leading to a sizeable increase in the branching ratios.
It is evident  that the branching ratios of both
channels could therefore increase to an level accessible
to LHC experiments. For instance,
taking $M_{T}=1200\,{\rm{GeV}}$ and $\sin\theta_{L}=0.025$,
the  $H\rightarrow b\bar{s}$ decay width is about $435\,{\rm{eV}}$,
translating to the branching ratio is $1.36\times10^{-4}$,
comparable to the LHC observation of
$H\rightarrow\mu^{+}\mu^{-}$~\cite{PDG2022Higgs}.

\begin{figure}
\begin{center}
\includegraphics[scale=1.00]{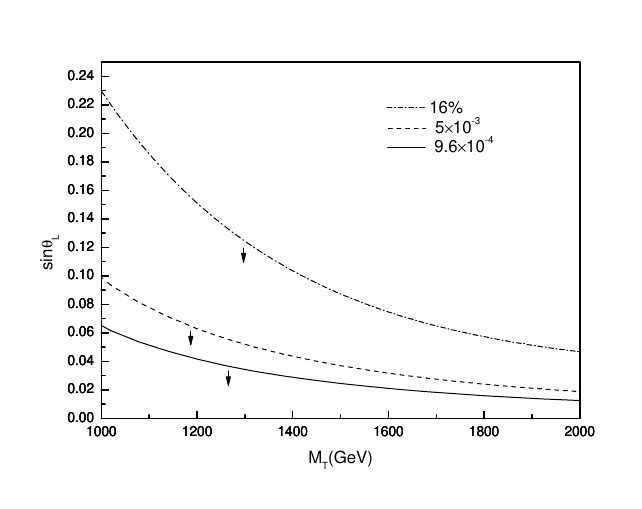}
\caption{The allowed region of $\{M_{T},\sin\theta_{L}\}$
parameter space determined from different
$H\rightarrow b\bar{s}$ branching ratios upper bounds. For each choice
of the branching ratio, the allowed area is below the line
as depicted by the arrow. }  \label{allowedarea}
\end{center}
\end{figure}

To compare the relative contribution of
each diagram in Fig.~\ref{toppartnercontributon},
we list the values
\begin{equation}
\mathcal{F}^{{\rm{VL}}}_{i}=4m_{b}m_{W}^{2}(\sqrt{2}G_{f})^{3/2}\mathcal{C}_{i},
\quad \quad i=1,\,2,\,...,\,5 \label{sizeoffivetypetoppartner}
\end{equation}
for $H\rightarrow b\bar{s}$ and for $H\rightarrow b\bar{d}$ for
$M_{T}=1500{\rm{GeV}},\,\sin\theta_{L}=0.02$ in
Table~\ref{coefficientsofhtobswithvectortop} and
Table~\ref{fourcoeffofhtobdwithvectortop}, respectively.
The results indicate that the third diagram
in Fig.~\ref{toppartnercontributon} is the dominant contribution
to the amplitude in Eq.~(\ref{totalamplitudebytoppartner}).

It is foreseeable that if we increase both $M_{T}$ and $\sin\theta_{L}$
simultaneously or solely one of them, the evaluated $H\rightarrow b\bar{s}$
width will exceed the total width of the Higgs boson.
For example, with $\sin\theta_{L}=0.08$,
the left panel of Fig.~\ref{varationoftheHtobsbarbranchfrac}
shows that the branching ratio exceeds $100\%$ for $M_{T}>1900{\rm{GeV}}$.
The situation is not physical and must be excluded.
In other words, the $H\rightarrow b\bar{s}$
branching fraction provides stringent constraints
on the $\{M_{T},\sin\theta_{L}\}$ parameter space.
The analysis in Ref.~\cite{Kamenik:2024mpe}
proposes three different upper bounds on the $H\rightarrow b\bar{s}$
branching ratios. The first one is from
Higgs boson studies at the LHC~\cite{ATLAS2021exp, Nature.607.60},
imposing an upper limit on undetermined
decays ${\rm{Br}}(H\rightarrow {\rm{undet}})<0.16$ at the $95\%$
confidence level (CL). The second is based on
a probabilistic model and if only the $b$-tagger
is used, the upper bound ${\rm{Br}}(H\rightarrow b\bar{s})<5\times10^{-3}$
is obtained at the $95\%$ CL. If both the $b$-tagger
and $s$-tagger are considered
in the probabilistic model,
the upper limit lowered to
${\rm{Br}}(H\rightarrow b\bar{s})<9.6\times10^{-4}$ at the $95\%$ CL.
Since at present there is no direct experimental
data on $H\rightarrow b\bar{s}$,
it is advisable to take the three upper limits
as inputs to constrain the allowed $\{M_{T},\sin\theta_{L}\}$ parameter
space as presented in Fig.~\ref{allowedarea}.
It is evident that a lower branching ratio bound leads
to a smaller area of allowed allowed $\{M_{T},\sin\theta_{L}\} $
parameter space. Since the upper limit $16\%$ incorporates
all the contributions from undetermined decays of the Higgs boson,
we conclude that if the mass of the vector-like top partner
is less than $2000\,{\rm{GeV}}$, then $\sin\theta_{L}$ is less than $0.24$.

An important issue we would like to address is
the tagging efficiencies in the observation of $H\rightarrow b\bar{s}$.
In this decay there are two jets in the final state.
To single out the $b\bar{s}$ final state,
$b$-tagging and $s$-tagging are indispensable.
We may follow the techniques of the probabilistic model
proposed in Ref.~\cite{Kamenik:2024mpe}.
Applying the $b$-tagger and $s$-tagger to the two jets,
the events are distributed in $(n_{b}, n_{s})\in\{(0,0),
(1,0), (0,1), (2,0), (1,1),(0,2)\}$ bins,
where $n_{b}$ and $n_{s}$ denotes the numbers of $b$-tagger
and $s$-tagger in the events, respectively.
The $b$-tagger and $s$-tagger efficiencies are denoted by the parameters
$\epsilon^{b}_{\beta}$ and $\epsilon^{s}_{\beta}$ where $\beta=\{g,s,c,b\}$
labels the flavor of the initial parton.
The signal of the $H\rightarrow b\bar{s}$ decay mostly
populates the $(1,1)$ bin while the other five bins
constrain the backgrounds. In order to scan over
possible tagger efficiencies, we
assume $\epsilon^{b}_{b}=\epsilon^{s}_{s}$~(true positive rate, TPR)
and $\epsilon^{b}_{gsc}=\epsilon^{s}_{gcb}$~(false positive rate, FPR).
Then the upper bounds of the $H\rightarrow b\bar{s}$
branching ratio can be expressed as function of TPR and FPR,
allowing the observation of $H\rightarrow b\bar{s}$ at the LHC.
By this approach, the working point (TPR, FPR)$=(0.80,0.004)$ leads to
the upper bound ${\rm{Br}}(H\rightarrow b\bar{s})<9.6\times10^{-4}$.
This bound is close to the branching ratio of $H\rightarrow \mu^{+}\mu^{-}$
observed by LHC. According to the correlations
between $H\rightarrow b\bar{s}$ and $B_{s}\rightarrow \mu^{+}\mu^{-}$
given in Ref.~\cite{Chiang:2017etj}, it is possible to detect
this decay at the LHC. On the other hand,
the $b$-tagging channel $H\rightarrow b\bar{b}$
has been observed at the LHC with the
branching ratio $53\%$~\cite{ATLAS:2017bic, CMS:2017odg},
it is the background for $H\rightarrow b\bar{s}$ and vice versa.
Therefore, if the $b$-tagger and $s$-tagger efficiencies can be increased,
isolating the signal of $H\rightarrow b\bar{s}$ is feasible.
However, considering the hadronic noise at the LHC,
the ILC~\cite{Bambade:2019fyw} may provide a better
environment for discovering $H\rightarrow b\bar{s}$.

\subsection{extension to $H\rightarrow\gamma\gamma, Z\gamma$
and two-doublet vector-like quark model }

Similar to $H\rightarrow b\bar{s}$, the two decays $H\rightarrow \gamma\gamma, Z\gamma$
are also affected by the presence of a singlet vector-like top partner
at leading order and have been investigated in
Refs.~\cite{Dawson:2012di, Arhrib:2016rlj, He:2020suf}.
The model in Ref.~\cite{Dawson:2012di} is in
line with our model in this work,
and the results show that the deviation of the following ratio
\begin{equation}
R=\frac{{\rm{Br(H\rightarrow\gamma\gamma)_{VL}}}}
{{\rm{Br(H\rightarrow\gamma\gamma)_{SM}}}},  \label{htotwogammasbranchingratio}
\end{equation}
from $1$ is always less than $1\%$ for small
mixing ($\sin\theta_{L}<0.20$) at $M_{T}=1\,{\rm{TeV}}$.
Within the framework of a type-\uppercase\expandafter{\romannumeral 2}
two-Higgs doublet model embedding the vector-like singlet top partner,
in Refs.~\cite{Arhrib:2016rlj} the two
processes $H\rightarrow \gamma \gamma, Z\gamma$ are investigated.
Results indicate that in order to compatible with the observed branching ratios of
these two decays, taking the value $M_{T}=1\,{\rm TeV}$,
the mixing angle satisfies $|\sin\theta_{L}|<0.25$.
While the analysis in Ref.~\cite{He:2020suf} is based on
a SM extension with a vector-like singlet top partner
plus a real singlet scalar $S$,
from the perturbative unitarity bounds on
the Yukawa coefficients $y^{tT}_{L,R}$,
the values $M_{T}=400\,{\rm GeV}$, $\sin\theta_{L}=0.20$
should be taken to get the optimal situation.
This implies that if the mass of the vector-like
partner significantly surpasses $1\,{\rm{TeV}}$
as the value employed in our analysis,
the resulting branching ratios may contradict
the experimental constraints. Therefore,
supposing the mass of the top partner
is above $1\,{\rm{TeV}}$, in order to guarantee
that the branching ratios of $H\rightarrow \gamma\gamma, Z\gamma$
are consistent with present experiments,
the value of $\sin\theta_{L}$ should be taken lower
than those obtained in Refs.~\cite{Arhrib:2016rlj, He:2020suf}.

In addition to the singlet quark model
employed in this work, there are vector-like
doublet and triplet models~\cite{Aguilar-Saavedra:2013qpa}, where
the new vector-like heavy quarks of these models can also contribute to
$H\rightarrow b\bar{s}, b\bar{d}$ and other
quark flavor-changing decays of the Higgs boson.
A full investigation of the effects of these models
on the quark flavor changing decays of the Higgs boson
will modify the main content of this work. Thus we take
the vector-like doublet model as an example and
briefly comment its implications
for the flavor changing decay of the Higgs boson.
Since the flavor changing neutral current are forbidden
at tree level in the SM, the singlet top partner
contributes to the process $H\rightarrow c\bar{u}$
at next-to-leading order. However, the situation will
be changed in the doublet vector-like quark model.
In this circumstance the Higgs boson will couple to
the down-type vector-like quark $B_{L}$ and $B_{R}$
via the following Lagrangian~\cite{Dawson:2012di}
\begin{equation}
\mathcal {L}\supset-c_{bb}\frac{m_{b}}{v}\bar{b}_{L}b_{R}h
-c_{BB}\frac{M_{B}}{v}\bar{B}_{L}B_{R}h-c_{bB}\frac{m_{b}}{v}\bar{b}_{L}B_{R}h
-c_{Bb}\frac{M_{B}}{v}\bar{B}_{L}b_{R}h+{\rm{H.c.}} \,,\label{twodoubletlagrangianofdowntype}
\end{equation}
where the definition of the parameters can be
found in Eq.~(51) of Ref.~\cite{Dawson:2012di}.
In this case we can explore the effects of
the vector-like down-type quark to the decay $H\rightarrow c\bar{u}$ at leading order.

\section{summary}
\label{summary_section}
The discovery of the SM-like Higgs boson opens a new window to
explore  quark flavor changing processes. In this paper, we
firstly presented a comprehensive analysis
of the $H\rightarrow b\bar{s}$ and $H\rightarrow b\bar{d}$ branching
ratios at leading order in the SM. The results
agree with the existing work obtained
in the SM~\cite{Benitez-Guzman:2015ana, Aranda:2020tqw, Farrera:2020bon}.

Subsequently, based on the vector-like singlet
top partner model, the $H\rightarrow b\bar{s}$
and $H\rightarrow b\bar{d}$ branching ratios of were evaluated.
Further results indicate that by tuning
the mass of the top partner and the mixing angle,
the branching ratios of both channels will increase significantly
to the level accessible to LHC experiments.
Then combining our results with three different
upper limits on the $H\rightarrow b\bar{s}$ branching ratio,
the allowed (two-dimensional) $M_{T}-\sin\theta_{L}$
parameter space was determined. According to our results,
assuming the top partner mass   is less than $2000\,{\rm{GeV}}$,
the mixing angle should satisfy $\sin\theta_{L}<0.24$.

Combining with a probabilistic model~\cite{Kamenik:2024mpe},
tagging efficiencies and detection feasibility of the $H\rightarrow b\bar{s}$
decay are carefully considered. Our analysis shows that it is
promising to detect $H\rightarrow b\bar{s}$ at the LHC,
but high statistics is needed.

Since we only consider the singlet top partner model,
the up-type flavor changing final state $c\bar{u}$
has not been taken into account, but can be
included in extensions to doublet or triplet vector-like models.
Such studies will be explored in our future work.

\section{acknowledgements}
TGS is grateful for research funding from
the Natural Sciences and Engineering Research Council of Canada (NSERC).

\begin{appendix}
\section{the running of the strong coupling constant}  \label{expressionofalphas}

The up-to-date results for the $\overline{{\rm MS}}$-scheme
strong coupling is~\cite{ParticleDataGroup:2022pth}
\begin{equation}
\alpha_{s}(M_{Z})=0.1179\pm0.0009, \label{uptodatestrongcoupling}
\end{equation}
and using the central value we obtain $\Lambda^{(5)}_{{\rm{QCD}}}=0.208{\rm{GeV}}$.
By employing this value, up to three-loop approximation in
QCD~\cite{Tarasov:1980au, Larin:1993tp, Chetyrkin:1997sg, Baikov:2016tgj},
the running of $\alpha_{s}$ at some energy sale $\mu$ can be determined
\begin{eqnarray}
\alpha_{s}(\mu)&=&\frac{4\pi}{\beta_{0}\ln(\mu^{2}/\Lambda^{2})}
\Big\{1-\frac{2\beta_{1}}{\beta_{0}^{2}}
\frac{\ln\ln(\mu^{2}/\Lambda^{2})}{\ln(\mu^{2}/\Lambda^{2})}
+\frac{4\beta_{1}^{2}}{\beta_{0}^{4}\ln^{2}(\mu^{2}/\Lambda^{2})}
\Big[\Big(\ln\ln\frac{\mu^{2}}{\Lambda^{2}}
-\frac{1}{2}\Big)^{2}\nonumber\\
&+&\frac{\beta_{2}\beta_{0}}{8\beta_{1}^{2}}
-\frac{5}{4}\Big]\Big\}, \label{runningofthestrongcoupling}
\end{eqnarray}
where the coefficients are given by
\begin{equation}
\beta_{0}=11-\frac{2}{3}N_{f},\quad \beta_{1}=51-\frac{19}{3}N_{f},
\quad \beta_{2}=2857-\frac{5033}{9}N_{f}+\frac{325}{27}N^{2}_{f}.  \label{betazerotobetatwo}
\end{equation}
with $N_{f}$ being the number of the active quarks below the scale $\mu$.
\section{the dilogarithms}
The dilogarithm is defined as~\cite{lewindilog}
\begin{equation}
{\rm{Li}}_{2}(x)=\sum_{n=1}^{+\infty}\frac{x^{2}}{n^{2}}
=-\int_{0}^{x}\frac{\ln(1-t)}{t}\D t,\quad |x|<1  , \label{definitionofthedilog}
\end{equation}
and an equivalent definition is
\begin{equation}
{\rm{Li}}_{2}(x)
=-\int^{1}_{0}\frac{\ln(1-xt)}{t}\D t,   \label{equivalentdefofdilog}
\end{equation}
where there is a branch cut from $1$ to $+\infty$,
\begin{equation}
{\rm{Li}}_{2}(x+i\varepsilon)={\rm{Re}}\,{\rm{Li}}_{2}(x)
+i\pi\,{\rm{sgn}}(\varepsilon)\Theta(x-1)\ln x,\quad\quad \varepsilon\rightarrow 0   \label{cutofthedilog}
\end{equation}
and the step function $\Theta(x)$ and the ${\rm{sgn}}(x)$ are as follows
\begin{equation}
\Theta(x)=\begin{cases}
1,\quad \quad    x>0\\
0, \quad\quad x<0
\end{cases}    \label{thetsepfunction}
\end{equation}
and
\begin{equation}
{\rm{sgn}}(x)=\begin{cases}
1,\quad \quad    x>0\\
-1, \quad\,\, x<0
\end{cases} .            \label{thesignfunction}
\end{equation}
Two other useful formulae are~\cite{lewindilog}
\begin{equation}
{\rm{Li}}_{2}(x,\theta)={\rm{Re}}\,{\rm{Li}}_{2}(xe^{i\theta})
=-\frac{1}{2}\int_{0}^{x}
\frac{\ln(1-2t\cos\theta+t^{2})}{t}\D t,   \label{secondformulaofdilog}
\end{equation}
and
\begin{equation}
{\rm{Li}}_{2}(x)+{\rm{Li}}_{2}(\frac{1}{x})
=\frac{\pi^{2}}{3}-\frac{1}{2}\ln^{2}(x)
-i\pi\ln x, \quad x>1.  \label{continuationofdilog}
\end{equation}

\section{Useful integrals in the evaluation} \label{frequentlyusedintegrals}

The first integral frequently used in our evaluation is
\begin{equation}
F(a,b,c)=\int_{0}^{1} \ln(ax^{2}+bx+c
-i\varepsilon)\D x,  \label{singlelogintegral}
\end{equation}
If $b^{2}-4ac>0$, there are two zeros of the argument in the range $[0,\,1]$,
the logarithm can develop imaginary part, and
the result is~\cite{npb323.267}
\begin{eqnarray}
F(a,b,c)&=&\ln(a-i\varepsilon)-2
+(1-x_{-})\ln(1-x_{-}+i\varepsilon)
+x_{-}\ln(-x_{-}+i\varepsilon)\nonumber\\
&+&(1-x_{+})\ln(1-x_{+}-i\varepsilon)
+x_{+}\ln(-x_{+}-i\varepsilon),   \label{secondtypetintcaseone}
\end{eqnarray}
where
\begin{equation}
x_{+}=\frac{1}{2a}(-b+\sqrt{b^{2}-4ac}),\quad
x_{-}=\frac{1}{2a}(-b-\sqrt{b^{2}-4ac}).   \label{twozerosoffirstintcaseone}
\end{equation}
If $b^{2}-4ac<0$, then the argument of the logarithm is always positive, and
the result reads
\begin{eqnarray}
F(a,b,c)&=&\ln(a+b+c-i\varepsilon)
-2+\frac{b}{2a}\ln\frac{a+b+c-i\varepsilon}{c}\nonumber\\
&+&\frac{\sqrt{4ac-b^{2}}}{a}
\Big(\arctan\frac{2a+b}{\sqrt{4ac-b^{2}}}
-\arctan\frac{b}{\sqrt{4ac-b^{2}}}\Big).   \label{secondtypeintcasetwo}
\end{eqnarray}

The second type integral is
\begin{equation}
G(\alpha;a,b,c)=\int_{0}^{1}
\frac{\ln(ax^{2}+bx+c-i\varepsilon)}{x+\alpha}\,\D x. \label{denifitionofsecodntype}
\end{equation}
In this case we should distinguish
between three cases: $\alpha>0$, $-1<\alpha<0$ and $\alpha<-1$.
If $b^{2}-4ac>0$, for the three cases of $\alpha$,
the result can be expressed uniformly as
\begin{eqnarray}
G(\alpha;a,b,c)
&=&\ln\big|1+\alpha\big|\ln(a+b+c-i\varepsilon)-\ln\big|\alpha\big|\ln(c-i\varepsilon)\nonumber\\
&+&\ln\big|1+\frac{1}{\alpha}\big|\ln\Big|\frac{a\alpha^{2}-b\alpha+c}{a+b+c}\Big|\nonumber\\
&+&{\rm{Li}}_{2}\Big[\frac{\alpha}{\alpha+x_{+}}-i\varepsilon\,{\rm{sgn}}(\alpha)\Big]
-{\rm{Li}}_{2}\Big[\frac{1+\alpha}{\alpha+x_{+}}-i\varepsilon\,{\rm{sgn}}(1+\alpha)\Big]\nonumber\\
&+&{\rm{Li}}_{2}\Big[\frac{\alpha}{\alpha+x_{-}}+i\varepsilon\,{\rm{sgn}}(\alpha)\Big]
-{\rm{Li}}_{2}\Big[\frac{1+\alpha}{\alpha+x_{-}}+i\varepsilon\,{\rm{sgn}}(1+\alpha)\Big].  \label{intsecodntypegreaterthanzero}
\end{eqnarray}
If $b^{2}-4ac<0$, the argument of the logarithm is always positive, and
by employing Eq.~(\ref{secondformulaofdilog}), we obtain
\begin{equation}
G(\alpha;a,b,c)=\ln D\ln\big|1+\frac{1}{\alpha}\big|
-2{\rm{Li}}_{2}\big(\frac{1+\alpha}{m},\theta\big)
+2{\rm{Li}}_{2}\big(\frac{\alpha}{m},\theta\big) \label{intsecodtypelessthanzero}
\end{equation}
where
\begin{eqnarray}
D&=&a\alpha^{2}-b\alpha+c,\quad m=\sqrt{\alpha^{2}
-\alpha\frac{b}{a}+\frac{c}{a}},\quad
\theta=\arccos\frac{2a\alpha-b}{2\sqrt{aD}}. \label{definitionofDinsecondtype}
\end{eqnarray}
For numerical convenience we recast Eq.~(\ref{intsecodtypelessthanzero}) into the
following form
\begin{equation}
G(a,b,c)=\ln D\ln\big|1+\frac{1}{\alpha}\big|
-2{\rm{Re}}\,{\rm{Li}}_{2}\big(x_{1}e^{i\theta}\big)
+2{\rm{Re}}\,{\rm{Li}}_{2}\big(x_{2}e^{i\theta}\big),    \label{realpartexpr4ssion}
\end{equation}
with
\begin{equation}
x_{1}=\frac{(1+\alpha)\sqrt{a}}{\sqrt{D}},\quad\quad
x_{2}=\frac{\alpha\sqrt{a}}{\sqrt{D}}. \label{twoparametersofdilog}
\end{equation}

\section{Passarino-Veltman functions}   \label{PVcoefficients}

The following two- and three-point functions are
frequently needed in the evaluation of the amplitudes.
By employing dimensional regularization,
setting $d=4-2\epsilon$, we obtain the two-point functions $B_{0}$ constrained by $p^{2}=0$
\begin{eqnarray}
\mathcal{B}_{0}&=&\int\frac{\D^{d}k}{(2\pi)^{d}}
\frac{1}{(k^{2}-m_{W}^{2}+i\varepsilon)[(k-p)^{2}-m^{2}_{q}+i\varepsilon]} \nonumber\\
&=&\frac{i}{(4\pi)^{2}}
\left[\frac{1}{\epsilon}-\gamma_{E}+\ln(4\pi)\right]+B_{0}(m_{q})
+\mathcal {O}(\epsilon),  \label{twopointzero}
\end{eqnarray}
where $\gamma_{E}=0.5772...$ is the Euler-Mascheroni constant,
and $B_{0}(m_{q})$ is given by
\begin{equation}
B_{0}(m_{q})=-\frac{i}{(4\pi)^{2}}
\Big(-1+\ln\frac{m_{W}^{2}}{\mu^{2}}
+\frac{m_{q}^{2}}{m_{W}^{2}-m_{q}^{2}}
\ln\frac{m_{W}^{2}}{m^{2}_{q}}\Big).  \label{twopointbzeromq}
\end{equation}
The vector two-point function is
\begin{eqnarray}
B_{\mu}&=&\int\frac{\D^{d}k}{(2\pi)^{d}}
\frac{k_{\mu}}{(k^{2}-m_{W}^{2}+i\varepsilon)
[(k-p)^{2}-m^{2}_{q}+i\varepsilon]}\nonumber\\
&=&\frac{ip_{\mu}}{32\pi^{2}}
\left[\frac{1}{\epsilon}-\gamma_{E}+\ln(4\pi)\right]
+B_{1}(m_{q})p_{\mu}
+\mathcal {O}(\epsilon),       \label{twopointmu}
\end{eqnarray}
with
\begin{equation}
B_{1}(m_{q})=-\frac{i}{32\pi^{2}}\Big[-\frac{3}{2}
+\ln\frac{m_{W}^{2}}{\mu^{2}}-\frac{m_{q}^{2}}{m_{W}^{2}-m_{q}^{2}}
+\frac{m_{q}^{2}(2m_{W}^{2}-m_{q}^{2})}{(m_{W}^{2}-m_{q}^{2})^{2}}
\ln\frac{m_{W}^{2}}{m_{q}^{2}}\Big].  \label{twopointbonemq}
\end{equation}
The general scalar and vector three-point functions are defined as
\begin{eqnarray}
C_{0}(m_{1},m_{2},m_{3},p_{1}^{2},p_{2}^{2})
&=&\int\frac{\D^{d}k}{(2\pi)^{d}}\frac{1}{D_{1}D_{2}D_{3}},\nonumber\\
C_{\mu}(m_{1},m_{2},m_{3},p_{1},p_{2})&=&\int\frac{\D^{d}k}{(2\pi)^{d}}\frac{k_{\mu}}
{D_{1}D_{2}D_{3}}
=C_{1}p_{1\mu}+C_{2}p_{2\mu},  \label{generalthrpointfunc}
\end{eqnarray}
where
\begin{eqnarray}
D_{1}&=&k^{2}-m_{1}^{2}+i\varepsilon,\nonumber\\
D_{2}&=&(p_{1}-k)^{2}-m_{2}^{2}+i\varepsilon,\nonumber\\
D_{3}&=&(p_{2}-k)^{2}-m_{3}^{2}+i\varepsilon.  \label{threegeneraldenominators}
\end{eqnarray}
In order to express the three-point
functions in a concise form, it is convenient
to define the following parameters formed by the masses in the evaluation
\begin{eqnarray}
&&a_{1}=m_{H}^{2},\quad b_{1}
=m_{1}^{2}-m_{H}^{2}-m_{2}^{2},\quad c_{1}=m_{2}^{2},\nonumber\\
&&a_{2}=m_{b}^{2},\quad \,\,b_{2}
=m_{1}^{2}-m_{b}^{2}-m_{3}^{2},
\quad\,\, \,c_{2}=m_{3}^{2},\nonumber\\
&&a_{3}=m_{1}^{2}-m_{3}^{2},\quad b_{3}=c_{2}=m_{3}^{2},\nonumber\\
&&\alpha=\frac{m_{2}^{2}-m_{3}^{2}}{m_{b}^{2}-m_{H}^{2}},
\quad \beta=\frac{m_{3}^{2}-m_{2}^{2}}{m_{H}^{2}}.  \label{quadraticparameters}
\end{eqnarray}
In the case $p_{1}^{2}=m_{H}^{2},\,p_{2}^{2}=m_{b}^{2}$, the results read
\begin{eqnarray}
C_{0}(m_{1},m_{2},m_{3},p_{1}^{2},p_{2}^{2})
&=&\frac{-i}{(4\pi)^{2}(m_{H}^{2}-m_{b}^{2})}\Big[G(a_{2},b_{2},c_{2})
-G(a_{1},b_{1},c_{1})\Big], \label{functionczeroscalar}\\
C_{1}(m_{1},m_{2},m_{3},p_{1}^{2},p_{2}^{2})
&=&\frac{-i}{(4\pi)^{2}(m_{H}^{2}-m_{b}^{2})}
\Big\{1+F(a_{1},b_{1},c_{1})\nonumber\\
&-&(1+\alpha)\Big[\ln\big|1+\frac{1}{\alpha}\big|+G(a_{1},b_{1},c_{1})\Big]\Big\}\nonumber\\
&-&\frac{i}{(4\pi)^{2}(m_{H}^{2}-m_{b}^{2})^{2}}
\Big\{\Big[\frac{c_{2}}{\alpha}\ln(c_{2}-i\varepsilon)
-\frac{a_{2}+b_{2}+c_{2}}{1+\alpha}\ln(a_{2}+b_{2}+c_{2}-i\varepsilon)\nonumber\\
&+&2a_{2}\Big(1+F(a_{2},b_{2},c_{2})\Big)
+(b_{2}-2\alpha a_{2})\Big(\ln\big|1+\frac{1}{\alpha}\big|
+G(a_{2},b_{2},c_{2})\Big)\Big]\nonumber\\
&-&(a_{2}\leftrightarrow a_{1},
b_{2}\leftrightarrow b_{1}, c_{2}\leftrightarrow c_{1})\Big\}
+\mathcal {O}(\epsilon),   \label{theconecoefficient}\\
C_{2}(m_{1},m_{2},m_{3},p_{1}^{2},p_{2}^{2})
&=&\frac{-i}{(4\pi)^{2}(m_{H}^{2}-m_{b}^{2})}
\Big\{-1-F(a_{2},b_{2},c_{2})\nonumber\\
&+&(1+\alpha)\Big[\ln\big|1+\frac{1}{\alpha}\big|
+G(a_{2},b_{2},c_{2})\Big]\Big\}\nonumber\\
&-&\frac{i}{(4\pi)^{2}(m_{H}^{2}-m_{b}^{2})^{2}}
\Big\{\Big[\frac{c_{1}}{\alpha}\ln(c_{1}-i\varepsilon)
-\frac{a_{1}+b_{1}+c_{1}}{1+\alpha}\ln(a_{1}+b_{1}
+c_{1}-i\varepsilon)\nonumber\\
&+&2a_{1}\Big(1+F(a_{1},b_{1},c_{1})\Big)
+(b_{1}-2\alpha a_{1})\Big(\ln\big|1+\frac{1}{\alpha}\big|
+G(a_{1},b_{1},c_{1})\Big)\Big]\nonumber\\
&-&(a_{1}\leftrightarrow a_{2},
b_{1}\leftrightarrow b_{2}, c_{1}\leftrightarrow c_{2})\Big\}
+\mathcal {O}(\epsilon).   \label{thectwocoefficient}
\end{eqnarray}

\end{appendix}

\end{document}